\documentclass[aps,showpacs,prep,graphics,twocolumn]{revtex4}
\usepackage{amssymb}
\usepackage[dvips]{graphicx}
\usepackage{caption}
\usepackage{amsmath}
\usepackage{bm}
\usepackage{epsfig}

\begin{document}

\title{ Higgs inflation in complex geometrical scalar-tensor theory of gravity.}

\author{ $^{1}$ Jos\'e Edgar Madriz Aguilar\thanks{E-mail address: madriz@mdp.edu.ar}, $^{2}$ J. Zamarripa, $^{1}$ M. Montes and $^{3}$ C. Romero} 
\affiliation{$^{1}$ Departamento de Matem\'aticas, Centro Universitario de Ciencias Exactas e ingenier\'{i}as (CUCEI),
Universidad de Guadalajara (UdG), Av. Revoluci\'on 1500 S.R. 44430, Guadalajara, Jalisco, M\'exico.\\
\\
$^{2}$ Centro Universitario de los Valles, Carretera Guadalajara-Ameca Km 45.5, C. P. 46600, Ameca, Jalisco, M\'exico.\\
and\\
$^{3}$ Departamento de F\'isica, Universidade Federal da Para\'iba, Caixa Postal 5008, 58059-970, Jo\~{a}o Pessoa, PB, Brazil.\\
E-mail: jose.madriz@academicos.udg.mx, zama\_92@live.com.mx, mariana.montes@academicos.udg.mx, cromero@fisica.ufpb.br }

\begin{abstract}
We derive a Higgs inflationary model in the context of a complex geometrical scalar-tensor theory of gravity. In this model the Higgs inflaton scalar field has geometrical origin playing the role of the Weyl scalar field in the original non-riemannian background geometry. The energy scale enough to generate inflation from the Higgs energy scale is achieved due to the compatibility of the theory with its background complex Weyl-integrable geometry. We found that for a number of e-foldings $N=63$, a nearly scale invariant spectrum for the inflaton is obtained with an spectral index $n_s\simeq 0.9735$ and a scalar to tensor ratio $r\simeq 0.01$, which are in agreement with Planck observational data.
\end{abstract}

\pacs{04.50. Kd, 04.20.Jb, 02.40k, 11.15 q, 98.80.Cq, 98.80}
\maketitle

\vskip .5cm
 Weyl-Integrable geometry, geometrical scalar-tensor gravity, Higgs cosmological inflation.

\section{Introduction}

Inflationary models represent a cornerstone of modern cosmology. By postulating the existence of the inflaton scalar field, inflation solves the old problems of the big bang cosmology and also provides a mechanism to explain the formation of cosmological structure. In this theory the inflaton must be capable to generate the enough vaccuum energy density to have a suitable model compatible with CMB observational data and the matter distribution in the universe. In the literature we can find   different inflationary models that use more than one scalar field, as for example the hybrid inflation models \cite{HYI1,HYI2,HYI3,HYI4}.  \\

  However, until now, the only scalar particle that has experimental evidence of his existence is the Higgs boson \cite{HI0A,HI0B}. The idea that the inflaton field might be the same as the Higgs scalar field has already been considered \cite{HI1}. The main problem of this idea relies in the fact that the energy scale of the Higgs field is too small to generate the enough quantity of inflation requiered to solve the problems of the big bang cosmology. In particular, to have the enough inflation to solve the big bang problems, the inflaton is estimated to have a mass $\sim 10^{13}$ {\it GeV}, and in some models it prefers a small-interacting  quartic coupling constant $\lambda\leq10^{-9}$ \cite{HI1,HI1A,HI1B}. However, all the parameters associated with the Higgs field are determined at {\it TeV} scale, such as the dimensionless Higgs quartic coupling $0.11<\lambda<0.27$ \cite{HI1B,HI1C}. Models attempting to solve this problem have already appeared in the literature, which in much are non-minimal coupling models \cite{HI2,HI3,HI4,HI5}.\\

On the other hand, scalar-tensor theories incorporate a scalar field in the action. However, for some researchers it is not so clear if the scalar field describes gravity or matter \cite{cont}. This happens in the so called Jordan frame. By means of a conformal transformation of the metric appears the Einstein frame. In the Jordan frame gravity exhibits a non-minimal coupling with the scalar field while in the Einstein frame it is obtained a minimal coupling \cite{c12}. The main controversy relies in determine which of the  both frames is the physical one. In the li\-terature we can find opinions in favor of one or the other \cite{cont}. However, on the other hand, it is a well-known fact that a geometry is characterized by the compatibility condition between the connection and the metric: $\nabla_{\mu} g_{\alpha\beta}=N_{\alpha\beta\mu}$. However, in general the compatibility condition does not remain invariant only under conformal transformations of the metric. Therefore, the usual manner in which we can  pass from the Jordan to Einstein frame in standard scalar-tensor theories, changes the background geometry, and this is why the physics in one or another frame can be  different. In particular geodesic observers in one frame are not in the other \cite{cont,c10,c13}.\\

 This controversy can be alleviated if the background geometry is not fixed {\it apriori} as Riemannian. This is the main idea in a recently introduced new kind of scalar-tensor theories known as geometrical scalar-tensor theories of gravity \cite{c10,c13}. 
 In this theories the background geometry is obtained via the Palatini variational principle. The resulting geometry is one of the Weyl-integrable type \cite{c10,c13}. As a consequence, the scalar field that appears in scalar-tensor theories becomes part of the affine structure of the space-time and in this sense can be considered as geometrical in origin. Hence, the background geometry is esentially the same for both the Weyl and the Riemann frames, which are the analogous for the Jordan and Einstein frames in usual scalar-tensor theories. Hence, the ambiguity about the nature of the scalar field that usually arises in standard scalar-tensor theories and the controversy between the two frames is not present in this new approach \cite{cont,c10,c13}.
 In the framework of this theory topics like $(2+1) $ gravity models, inflatio\-nary cosmology and cosmic magnetic fields, quintessence and some cosmological models  have been studied \cite{c14,H16,H17,H18}.\\

In this letter we extend the formalism of previous geome\-trical scalar-tensor theories to construct a geome\-trical Higgs inflationary model. The letter is organized as follows. Section I is left for a little motivation and introduction. In section II is developed the general formalism in the Weyl frame. In section III it is obtained the effective field action in the Riemann frame. In section IV we present a Higgs inflationary model. Finally,  section V is devoted to some final comments.

\section {Basic formalism in the Weyl frame}

 Let us start considering an action for a complex  scalar-tensor theory of gravity, which in vacuum is given by
 \begin{small}
 \begin{equation}\label{f1}
 {\cal S}=\frac{1}{16\pi}\int d^4x \sqrt{-g}\left[\tilde{\Phi}\tilde{\Phi}^\dagger {\cal R}+\frac{\tilde{W}(\tilde{\Phi}\tilde{\Phi} ^\dagger)}{\tilde{\Phi}\tilde{\Phi} ^\dagger}g^{\mu\nu}\tilde{\Phi}_{,\mu}\tilde{\Phi}_{,\nu} -\tilde{U}(\tilde{\Phi}\tilde{\Phi} ^\dagger)  \right] 
 \end{equation}
\end{small}
where ${\cal R}$ denotes the Ricci scalar, $ \tilde{W}(\tilde{\Phi}\tilde{\Phi} ^\dagger)$ is a well-behaved differentiable function of $\tilde{\Phi}\tilde{\Phi} ^\dagger$, the dagger $\dagger$ denotes transposed complex conjugate  and $\tilde{U}(\tilde{\Phi}\tilde{\Phi} ^\dagger)$ is a scalar potential. With the help of the transformation $\tilde{\Phi}=\frac{1}{\sqrt{G}}\Phi$ the action \eqref{f1} can be written in the form
 \begin{equation}\label{ff1}
{\cal S}=\int d^4x \sqrt{-g}\left[\frac{\Phi\Phi^\dagger {\cal R}}{16\pi G}+\frac{\tilde{\omega}(\Phi\Phi ^\dagger)}{\Phi\Phi ^\dagger}g^{\mu\nu}\Phi_{,\mu}\Phi_{,\nu} -\tilde{V}(\Phi\Phi ^\dagger)  \right],
\end{equation}
where $\tilde{\omega}(\Phi\Phi^{\dagger})=\tilde{W}(\tilde{\Phi}\tilde{\Phi}^{\dagger})/(16\pi)$ and the redefined potential is $\tilde{V}(\Phi\Phi^{\dagger})=\tilde{U}(\tilde{\Phi}\tilde{\Phi}^{\dagger})/(16\pi)$ .
A Palatini variation of the action (\ref{ff1}) with respect to the affine connection leaves to the compatibility condition 
\begin{equation}\label{for1}
\nabla_{\mu}g_{\alpha\beta}=-[\ln (\Phi\Phi^{\dagger})]_{,\mu}g_{\alpha\beta}.
\end{equation}
Hence, the natural background geometry associated to (\ref{f1}) is a non-Riemannian geometry with a quadratic in $\Phi$ non-metricity and null torsion. However, through the field transformation $ \Phi = e^{-\varphi}$, the non-metricity in (\ref{for1}) can be linearized and written in the form
\begin{equation}\label{for2}
\nabla_{\mu} g_{\alpha\beta}=(\varphi + \varphi^{\dagger})_{,\mu}g_{\alpha\beta}.
\end{equation}
Notice that this compatibility condition is of the Weyl-Integrable  type.
Thus, in terms of the new field $\varphi$ the action (\ref{ff1}) reads
\begin{eqnarray}
{\cal S}=\int d^{4}x\sqrt{-g}e^{-(\varphi+\varphi^\dagger)}\left[\frac{\mathcal{R}}{16\pi G} + \hat{\omega}(\varphi+\varphi^\dagger)g^{\mu\nu}\varphi_{,\mu}\varphi^\dagger_{,\nu}\right.\nonumber\\
\label{f2}
\left.- \hat{V}(\varphi+\varphi^\dagger)\right],
\end{eqnarray}
where we have made the identifications  $\hat{\omega}(\varphi+\varphi^\dagger)=\tilde{\omega}(\varphi+\varphi^\dagger)e^{\varphi+\varphi^\dagger}$ and $ \hat{V}(\varphi+\varphi^\dagger)=\tilde{V}(\varphi+\varphi^\dagger)e^{\varphi+\varphi^\dagger}$. Now, we must note that the compatibility condition (\ref{for2}) remains invariant when we apply, at the same time, the transformations
 \begin{eqnarray}\label{f4}
  \bar{g}_{\mu\nu}&=& e^{f+f^\dagger}g_{\mu\nu},\\
  \label{f5}
    \bar{\varphi}&=& \varphi +f,\\
    \label{f55}
     \bar{\varphi}^\dagger&=& \varphi^{\dagger} +f^\dagger,
  \end{eqnarray}
where $f=f(x^{\alpha})$ is a well defined complex function of the space-time coordinates.   Thus for the action (\ref{f2}) to be an scalar under the group of transformations of the background geometry, it must be an inva\-riant under the diffeo\-morfism group and the  transformations (\ref{f4})-(\ref{f55}). However, under (\ref{f4}), (\ref{f5}) and (\ref{f55}) the kinetic term in (\ref{f2}) transforms as 
\begin{eqnarray}
 &&\sqrt{-\bar{g}}\, \bar{\hat{\omega}}(\bar{\varphi}+\bar{\varphi}^\dagger)\bar{g}^{\mu\nu}\bar{\varphi}_{,\mu}\varphi^\dagger_{,\nu}=\nonumber\\
 &&
 e^{2(f+f^\dagger)} \sqrt{-g}\, \hat{\omega} (\varphi +f+\varphi ^\dagger+f^\dagger)g^{\mu\nu}(\varphi_{,\mu}+f_{,\mu})(\varphi_{,\nu}^\dagger\nonumber\\
 \label{f6}
 &&
 +f^\dagger_{,\nu}),
\end{eqnarray}
which indicates that the kinetic term in \eqref{f2} results to be not invariant and consequently the action (\ref{f2}) is not either. In order to solve this problem we propose the new action
\begin{eqnarray}
&&{\cal S}=\int d^4x\sqrt{-g}\,e^{-(\varphi+\varphi^\dagger)}\left[\frac{\mathcal{R}}{16\pi G}+\widehat{\omega}(\varphi+\varphi^\dagger) g^{\mu\nu}\varphi_{:\mu}\varphi^\dagger_{:\nu}\right.\nonumber\\
\label{f9}
&&\left.
-e^{-(\varphi+\varphi\dagger)}\widehat{V}(\varphi+\varphi^\dagger)\right],
\end{eqnarray}
where we have introduced a gauge covariant derivative defined by $\varphi_{:\mu}=\,^{(w)}\nabla_{\mu}\varphi+\gamma B_{\mu}\varphi$, with  $B_{\mu}$ being a gauge vector field, $^{(w)}\nabla_{\mu}$ being the Weyl covariant derivative determined by (\ref{for2}) and  $\gamma$ is a pure imagi\-nary  coupling constant introduced to have the correct physical units. Thus, it is not difficult to verify that the invariance under (\ref{f4}) to (\ref{f55}) of (\ref{f9}) is achieved when the vector field $B_{\mu}$, the function $\hat{\omega}$ and the scalar potential $\hat{V}(\varphi)$, obey respectively the transformation rules
\begin{eqnarray}\label{f10a}
\bar{\varphi}\bar{B}_{\mu} &=& \varphi B_{\mu}-\gamma^{-1}f_{,\mu},\\
\label{f10aa}
\bar{\varphi}^\dagger\bar{B}_{\mu} &=& \varphi B_{\mu}+\gamma^{-1}f_{,\mu}^\dagger,\\
\bar{\hat{\omega}}(\bar{\varphi}+\bar{\varphi}^\dagger)&\equiv&\hat{\omega}(\bar{\varphi}+\bar{\varphi}^\dagger-f-f^\dagger)=\hat{\omega}(\varphi+\varphi^{\dagger}),\label{f10b}\\
\bar{V}(\varphi+\varphi^\dagger)&\equiv& V(\bar{\varphi}+\bar{\varphi}^\dagger-f-f^\dagger)=V(\varphi+\varphi^\dagger).\label{f10c}
\end{eqnarray}
Notice that \eqref{f10a} and (\ref{f10aa}) are transformation rules for the product $\varphi B_{\mu}$. Besides they have the same algebraic form of the algebra of the $U(1)$ group, used to describe the electromagnetic interaction. Thus, we may include a dynamics for $\varphi B_{\alpha}$ extending the action (\ref{f9}) by adding an electromagnetic type term in the form
\begin{small}
\begin{eqnarray}
&& {\cal S}=\int d^{4}x\sqrt{-g}\,e^{-(\varphi+\varphi^{\dagger})}\left[\frac{{\cal R}}{16\pi G}+\frac{1}{2}\hat{\omega}(\varphi+\varphi^{\dagger})g^{\alpha\beta}\varphi_{:\alpha}\varphi_{:\beta}\right.\nonumber\\
&&\left.
-e^{-(\varphi+\varphi^{\dagger})}\hat{V}(\varphi+\varphi^{\dagger})-\frac{1}{4}e^{(\varphi+\varphi^\dagger)}H_{\alpha\beta}H^{\alpha\beta}\right],
\label{f14}
\end{eqnarray}
\end{small}
where  $H_{\alpha\beta}=(\varphi B_{\beta})_{,\alpha}-(\varphi B_{\alpha})_{,\beta}$ is the field strength asso\-ciated to the gauge boson field $B_{\mu}$. The action (\ref{f14}) is an invariant action compatible with its background geo\-metry and originates a new kind of complex scalar-tensor theory of gravity. Given that its background geometry has a  non-metricity of the Weyl-Integrable type, we will refer to $(M,g,\varphi,\varphi^{\dagger},B_{\mu})$ as the Weyl frame. In this frame the dynamics is go\-ver\-ned by the field equations derived from the action (\ref{f14}). In addition, the transformations \eqref{f4} to \eqref{f55} can be interpreted geometrically as they lead from one frame $(M,g,\varphi,\varphi^{\dagger},B_{\mu})$ to another $(M,\bar{g},\bar{\varphi},\bar{\varphi}^{\dagger},\bar{B}_{\mu})$ sharing the same geo\-me\-try, the one  determined by 
\eqref{for2}. In this sense all the Weyl frames belong to the same equivalence class. However, there is one element of the class in which by redefining the metric tensor, an effective Riemannian geometry can be obtained. This issue will be the start point of the next section.

\section{Field equations in the Riemann frame}

 As it was mentioned in the previous section, the transformations (\ref{f4}), (\ref{f5}) and (\ref{f55}) lead from one Weyl frame $(M,g,\varphi,\varphi^\dagger,B_{\alpha})$ to another $(M,\bar{g},\bar{\varphi},\bar{\varphi}^\dagger,\bar{B}_{\alpha})$. However, for the particular choice $f=-\varphi$, we can define the effective metric $h_{\mu\nu}=\bar{g}_{\mu\nu}=e^{f+f^\dagger}g_{\mu\nu}$ such that $\bar{\varphi}=\bar{\varphi}^\dagger=0$. The interesting of this election is that in this case the condition (\ref{for2}) reduces to the effective Riemannian metri\-city condition: $\nabla_{\lambda} h_{\alpha\beta}=0$. For this reason we will refer to this frame $(M,\bar{g},\bar{\varphi}=0,\bar{\varphi}^\dagger=0,\bar{B}_{\alpha})=(M,h,\bar{B}_{\alpha})$, as the Riemann frame. We will use this terminology to differentiate it from the traditional Einstein and  Jordan frames employed in the scalar tensor theories we can find in the literature. The main reason to differentiate both terminologies is that in the traditional approaches the geodesics are not preserved under conformal transformations, while in the new kind of theories the geodesics are Weyl invariant \cite{c14}.\\ 
 
 In the Weyl frame the scalar field plays the role of a dilatonic geometrical scalar field while in the Riemann frame the Weyl field is no longer part of the affine structure. It means that when we go from the Weyl to the Riemann frame, the Weyl field pass from being  geometri\-cal to a physical one. In addition, once we are in the Riemann frame the action needs to be invariant only under the diffeomorphism group, and it implies that the geometrical invariance requirement for the gauge vector field $B_{\mu}$ given by (\ref{f10a}) is no more valid in this  frame. Thus due to the change of geometry, the scalar field $\varphi$ and the gauge vector field $B_{\mu}$ have different properties and  interpretations in each frame. \\

Once we have established some of the physical and geome\-trical differences between both frames, it is not diffi\-cult to verify that the action (\ref{f14}) in the Riemann frame acquires the form
\begin{eqnarray}
{\cal S}=\int d^{4}x\sqrt{-h}\left[\frac{\mathcal{R}}{16\pi G} +\hat{\omega} (\varphi+\varphi^\dagger) h^{\mu\nu}\mathbb{D}_\mu\varphi\mathbb{D}_\nu\varphi ^\dagger\right.\nonumber\\
\label{Rie2}
\left.-\widehat{V}(\varphi+\varphi^\dagger)-\frac{1}{4}H_{\mu\nu}H^{\mu\nu}\right],
\end{eqnarray}
where now the gauge covariant derivative becomes  $\mathbb{D}_{\mu}= \,^{(R)}\!\nabla_{\mu}+\gamma B_{\mu}$  and  the operator $^{(R)}\!\nabla_\mu$  denotes the Riemannian covariant derivative.\\

Thus, in order to restore the quadratic dependence in the scalar field, lost when we linearized \eqref{for1} to obtain \eqref{for2}, we introduce the field  transformations 
\begin{eqnarray}
\zeta &=&\sqrt{\xi}\,e^{- \varphi},\\
A_\mu&=&B_\mu \ln (\zeta/\sqrt{\xi}),
\end{eqnarray}
where $\xi$ is a constant introduced so that the field $\zeta$ has the correct physical units. Hence, the action (\ref{Rie2}) can be written as
\begin{eqnarray}
&& {\cal S}=\int d^{4}x\sqrt{-h}\left[\frac{\mathcal{R}}{16\pi G} +\frac{1}{2}\omega(\zeta\zeta ^\dagger) h^{\mu\nu} D _\mu \zeta( D _\nu \zeta) ^\dagger\right.\nonumber\\
\label{yq3}
&&
 \left. -V(\zeta\zeta ^\dagger)-\frac{1}{4}F_{\mu\nu} F^{\mu\nu}\right],
\end{eqnarray}
being $\mathcal{D}_\mu \zeta\equiv\zeta\mathbb{D}_\mu(\ln\frac{\zeta}{\sqrt{\xi}})=\,^{(R)}\nabla_{\mu}\zeta +\gamma A_{\mu}\zeta$ the effective covariant derivative, $F_{\mu\nu}\equiv\partial_{\mu}A_{\nu}-\partial_{\nu}A_{\mu}=-H_{\mu\nu}$ is the Faraday tensor and  where we have made the fo\-llo\-wing identifications
 	\begin{eqnarray}
 	 \frac{\omega(\zeta \zeta^\dagger)}{2}&\equiv& \frac{\widehat{\omega}(\ln\frac{\zeta \zeta ^\dagger}{\xi})}{\zeta \zeta^\dagger},\label{5.46} \\
 	V(\zeta\zeta^\dagger)&\equiv &\widehat{V}\bigg(\ln\frac{\zeta\zeta^\dagger}{\xi}\bigg).\label{5.45}
 \end{eqnarray}
  Notice that the action (\ref{yq3}) is invariant under the gauge transformations
 \begin{eqnarray}\label{5.48}
 \bar{\zeta}&=& \zeta e^{\gamma\theta(x)}\\
 \label{5.49}
 \bar{A}_\mu &=& A_\mu -\theta_{,\mu},
 \end{eqnarray}
 where $\theta(x)$ is a well-behaved function. Hence, due to the presence of the last term in \eqref{yq3} and the transformations \eqref{5.48} and (\ref{5.49}), we can interpret that $A_\mu$ can play the role of an electromagnetic potential.\\
 
 The action (\ref{yq3}) corresponds to an action of a complex scalar field minimally coupled to gravity in the presence of a free electromagnetic field where the scalar field has $U(1)$ symmetry.  Thus, due to the fact that the electromagnetic potential $A_{\mu}$ enters in the covariant derivative ${\cal D}_{\mu}$, the theory derived from (\ref{yq3}) can be interpreted as a gravitoelectromagnetic theory. Notice that in our formalism, the part of \eqref{yq3} that we relate with electromagnetism has its origin in the Weyl invariance of the action \eqref{f9}, which is not the case when electromagnetism is introduced in traditional approaches of scalar-tensor theories of gravity.
 
 \section{A Higgs inflation model}
 
 In this section we formulate a Higgs inflationary model from the gravitoelectromagnetic theory developed in the previous sections.  With this idea in mind let us consider the Higgs potential in the Weyl frame in the form
 \begin{equation}\label{hpot1}
 \tilde{V}(\Phi\Phi^{\dagger})=\frac{\lambda}{4}\left(\Phi\Phi^{\dagger}-\sigma^2\right)^2,
 \end{equation}
 where $\lambda=0.129$ and the vacuum expectation value for electroweak interaction $\sigma=246\,GeV$. These va\-lues accor\-ding to the best-fit experimental data \cite{expdata1,expdata2}. Thus, the Higgs potential in terms of the field $\zeta$ in the Riemann frame reads
  \begin{equation}\label{5.50}
  V(\zeta\zeta^\dagger)=\frac{\lambda}{4}\left(\frac{\zeta\zeta^\dagger}{\xi}-\sigma^2\right)^2.
  \end{equation}
 The minimum of the potential $||\zeta\zeta^\dagger||=\sqrt{\xi}\sigma$ results to be also inva\-riant under  \eqref{5.48}.  Howe\-ver, if we propose $\zeta=\zeta^\dagger$ we get $||\overline\zeta^2||\neq||\zeta^2|$, breaking in this manner the symmetry. Thus, excitations about the ground state of (\ref{5.50}) can be written in the form
 \begin{equation}\label{phiexp}
 \zeta(x^\mu)=\sqrt{\xi}\,\sigma+{\cal{Q}}(x^\mu),
 \end{equation}
 where ${\cal Q}(x)$ is the Higgs scalar field. 
 It can be verified with the help of \eqref{phiexp} that the kinetic term in (\ref{yq3}) gives
 \begin{eqnarray}
 \frac{\omega(\zeta)}{2}\mathcal D^\nu\zeta\mathcal{D}_\nu\zeta=\frac{\omega_{eff} ({\cal Q})}{2} \left(\partial^ {\nu} {\cal Q}\partial_{\nu}{\cal Q}-\gamma^2\xi\sigma^2A^\nu A_\nu\right.\nonumber\\
 \label{kt1}
 \left.
 -2\gamma^2\sqrt{\xi}\,\sigma {\cal Q} A ^\nu A_\nu-\gamma^2 {\cal Q}^2A^\nu A_\nu\right),
 \end{eqnarray}
 where $\omega_{eff}({\cal Q})=\omega(\sqrt{\xi}\sigma+{\cal Q})$.
 However, in order to develop a Higgs inflationary model, the cosmological principle restricts the existence of the field $A_{\mu}$ on large cosmological scales. Thus, it results convenient the gauge election: $\theta_{,\mu}=A_{\mu}$ or equivalently $\overline{A}_\mu=0$.  Under this gauge election, the terms in \eqref{kt1} that depend of the electromagnetic field $A_{\mu}$ become null and thus the action \eqref{yq3} leads to
 \begin{equation}\label{newac}
 {\cal S}=\int d^{4}x\sqrt{-h}\left[\frac{R}{16\pi G}+\frac{1}{2}\omega_{eff}({\cal Q})h^{\mu\nu}{\cal Q}_{,\mu}{\cal Q}_{\nu}-V_{eff}({\cal Q})\right],
 \end{equation}
 where $V_{eff}({\cal Q})=V(\sqrt{\xi}\sigma+{\cal Q})$. Now, in order to have a scalar field with a canonical kinetic term we use the field transformation
 \begin{equation}\label{ft}
 \phi(x^{\sigma})=\int\sqrt{\omega_{eff}({\cal Q})}\,d{\cal Q}.
 \end{equation}
Thus, the action for the Higgs field \eqref{newac} yields
\begin{equation}\label{yq4}
{\cal S}=\int d^{4}x\sqrt{-h}\left[\frac{\mathcal{R}}{16\pi G} +\frac{1}{2} h^{\mu\nu} \phi_{,\mu}\phi_{,\nu}-U(\phi)\right],
\end{equation}
where 
\begin{equation}\label{newpot}
U(\phi)=V_{eff}[{\cal Q}(\phi)]=\frac{\lambda}{4}\left[\frac{(\sqrt{\xi}\sigma+{\cal Q}(\phi))^2}{\xi}-\sigma^2\right]^2,
\end{equation}
is the potential written in term of the new field $\phi$.
Straithforward calculations show that the action (\ref{yq4}) leads to the field equations
  \begin{eqnarray}
    G_{\alpha\beta}=-8\pi G[\phi_{,\alpha}\phi_{,\beta}-\frac{1}{2}h_{\alpha\beta}\left(\phi^{,\mu}\phi_{,\mu} +2U(\phi)\right)], \label{Rie6}\\
   \Box\phi+U'(\phi)=0,\label{Rie7}
 \end{eqnarray}
 with $\Box$ denoting the D'Alambertian operator and the prime representing derivative with respect to $\phi$. Now, we consider a 3D-spatially flat Friedmann-Robertson-Walker metric in the form
\begin{equation}\label{inff7}
ds^2=dt^2-a^2(t)(dx^2+dy^2+dz^2),
\end{equation}
with $a(t)$ being the usual cosmological scale factor. As it is usually done in inflationary frameworks, the cosmologi-cal principle allow us to assume that the inflaton scalar field $\phi$, given by \eqref{ft}, can be written in the form
\begin{equation}\label{inff1}
\phi(x^{\lambda})=\phi_{c}(t)+\delta\phi(x^{\lambda}),
\end{equation}
where $\phi_{c}(t)=\left<\phi(x^{\lambda})\right>$, $\left<\delta\phi\right>=\left<\right.\delta\dot{\phi}\left.\right>=0$. Here $\delta\phi$ denotes the quantum fluctuations of the inflaton scalar field and $< >$ represents espectation value. It follows from the equations (\ref{Rie7}) and (\ref{inff7}) that the classical and quantum parts for the inflaton field can be written respectively as
\begin{eqnarray}
&& \ddot{\phi}_c+3H\dot{\phi}_c+U^{\prime}(\phi_c)=0,\label{inff2}\\
&& \ddot{\delta\phi}+3H\dot{\delta\phi}-\frac{1}{a^2}\nabla\delta\phi+U^{\prime\prime}(\phi_c)\delta\phi=0,
\label{inff3}
\end{eqnarray}
where $H=\dot{a}/a$ is the Hubble parameter. Now, conside\-ring that the universe is filled with a perfect fluid, the classical part of (\ref{Rie6}) leads to the Friedmann equations
\begin{eqnarray}\label{inff6}
&&H^2=\frac{\rho}{3M_p^2},\\
\label{inff8}
&&\dot{H}=-\frac{1}{2M_p^2}(\rho+p),
\end{eqnarray}
where $M_p=(8\pi G)^{-1/2}=2.45\cdot10^{18}\,GeV$ is our planckian mass convention,  $\rho=\frac{1}{2}\dot{\phi}_c^2+U(\phi_c)$ is the energy density and $p=\frac{1}{2}\dot{\phi}_c^2-U(\phi_c)$ is the pressure, all measured respect to a class of comoving observers. Under the slow-roll condition $|\dot{\phi}_c^2/2|\ll |U(\phi_c)|$, the equation of state parameter become $\eta_{\phi}=p/\rho\simeq-1$, which is a necessary condition to have inflation. In this manner, the classical part of the inflaton field is given by the equations (\ref{inff2}), (\ref{inff6}) and (\ref{inff8}), whereas their quantum fluctuations are governed by the expression (\ref{inff3}). \\

Now, by means of \eqref{inff2} and \eqref{inff6}, the classical part of the inflaton field $\phi_c$ is determined by
\begin{equation}\label{ulca1}
\dot{\phi_c}=-\frac{M_{p}}{\sqrt{3}}\frac{U^{\prime}(\phi_c)}{\sqrt{U(\phi_c)}}.
\end{equation}
Thus, in order to illustrate how the formalism works let us consider the anzats 
\begin{equation}\label{ulca2}
\omega_{eff}(Q)=\frac{1}{\left[1-\beta^2(\sqrt{\xi}\sigma+Q)^4\right]^{5/2}},
\end{equation}
where $\beta$ is a constant parameter with units of $M_{p}^{-2}$. Thus the equation \eqref{ft} yields
\begin{equation}\label{ulca3}
\phi=\frac{\sqrt{\xi}\sigma+Q}{\left[1-\beta^2(\sqrt{\xi}\sigma + Q)^4\right]^{1/4}}.
\end{equation} 
Therefore the potential \eqref{newpot} reads
\begin{equation}\label{ulca4}
U(\phi)=\frac{\lambda}{4\xi^2}\left(\frac{\phi^4}{1+\beta^2\phi^4}\right).
\end{equation}
After inflation begins when the condition $\beta\phi^4\ll 1$ holds, the potential \eqref{ulca4} becomes
\begin{equation}\label{potre}
U(\phi)\simeq \frac{\lambda}{4\xi^2}\phi^4.
\end{equation}
Thus, it follows from\eqref{ulca4} and \eqref{ulca1} that $\phi_c$ is given implicitly by 
\begin{eqnarray}
&&t-t_0+\frac{\beta^2}{6\alpha}\left(\phi_c^4\sqrt{1+\beta^2\phi_c^4}-\phi_0^4\sqrt{1+\beta^2\phi_0^4}\,\right)+\nonumber\\
&& \frac{2}{3\alpha}\left(\sqrt{1+\beta^2\phi_c^4}-\sqrt{1+\beta^2\phi_0^4}\,\right)+\nonumber\\
&& \frac{1}{2\alpha}tanh^{-1}\left(\frac{1}{\sqrt{1+\beta^2\phi_c^4}}\right)-\frac{1}{2\alpha}tanh^{-1}\left(\frac{1}{\sqrt{1+\beta^2\phi_0^4}}\right)\nonumber\\
&& =0,
\label{ulca5}
\end{eqnarray}
where $\phi_0=\phi(t_0)$ with $t_0$ being the time when inflation begins. Thus, in view that the number of e-foldings is given by
\begin{equation}\label{efol}
N(\phi)=M_p^{-2}\int_{\phi_e}^{\phi}\frac{U(\phi)}{U^{\prime}(\phi)}d\phi,
\end{equation}
being $\phi_e$ the value of the inflaton field at the end of inflation, the classical scalar field $\phi_{c}$ in terms of $N$ has the form
\begin{equation}\label{phiefol}
\phi_c(N)=\frac{\sqrt{\beta \Delta^{1/3}\left(\Delta^{2/3}-1\right)}}{\beta \Delta^{1/3}},
\end{equation}
where
\begin{equation}\label{phiefol1}
\Delta=12\beta NM_{p}^2+\sqrt{1+144N^2\beta^2M_{p}^4}.
\end{equation}
For the potential \eqref{potre} the expression \eqref{ulca5} reduces to 
\begin{equation}\label{sms1}
\phi_c(t)=\phi_e e^{2M_p\sqrt{\frac{\lambda}{3\xi^2}}\,(t_e-t)},
\end{equation}
which near to the end of inflation can be approximated by
\begin{equation}\label{sms2}
\phi_c(t)\simeq \phi_e\left[1+2M_p\sqrt{\frac{\lambda}{3\xi^2}}\left(t_e-t\right)\right],
\end{equation}
with $t_e$ denoting the time when inflation ends. Thus, employing \eqref{inff6}, \eqref{potre} and \eqref{sms1} it is obtained an approximated scale factor of the form
\begin{equation}\label{sms3}
a=a_e\exp \left[\frac{\phi_e^2}{8M_p^2}\left(1-\exp\left(4M_p\sqrt{\frac{\lambda}{3\xi^2}}\,(t_e-t)\right)\right)\right],
\end{equation}
where $a_e=a(t_e)$. For $t\simeq t_e$ \eqref{sms3} can be approximated by
\begin{equation}\label{sms4}
a(t)\simeq \tilde{a}_e\exp\left(\frac{\phi_e^2}{2M_p}\sqrt{\frac{\lambda}{3\xi^2}}\,t\right)
\end{equation}
where $\tilde{a}_e=a_e\exp\left(-\frac{\phi_e^2}{2M_p}\sqrt{\frac{\lambda}{3\xi^2}}\,t_e\right)$. Thus  the Hubble parameter associated with \eqref{sms3} is then
\begin{equation}\label{sms5}
H(t)=\frac{1}{\sqrt{3}M_p}\sqrt{\frac{\lambda}{4\xi^2}}\,\phi_e^2\exp\left(4M_p\sqrt{\frac{\lambda}{3\xi^2}}\,(t_e-t)\right).
\end{equation}
Therefore, near to the end of inflation \eqref{sms5} can be approxi\-mated by
\begin{equation}\label{sms6}
H(t)\simeq\frac{\phi_e^2}{\sqrt{3}M_p}\sqrt{\frac{\lambda}{4\xi^2}}\,\left[1+4M_p\sqrt{\frac{\lambda}{3\xi^2}}\,(t_e-t)\right].
\end{equation}
On the other hand, in order to have agreement with PLANCK data, Higgs inflation requires an energy scale corresponding to an initial Hubble parameter of the order $H_0\simeq 10^{11}-10^{12}\,GeV$, for an average  Higgs mass of the order $M_h\simeq125.7\,GeV$ \cite{Hmass1,Hmass2}. Therefore we obtain
\begin{equation}\label{chp1}
H_0\simeq\frac{\lambda}{2\sqrt{3}}\frac{1}{\beta\xi M_{p}}\simeq 10^{11}-10^{12}\,GeV.
\end{equation}
Using $\lambda=0.13$ and $M_p=1.22\cdot10^{19}\,GeV$ \cite{Hmass2},  we obtain that $\xi$ must vary in the interval: $\left[3.7528\cdot 10^{-14},3.7528\cdot 10^{-13}\right](\beta M_{p})^{-1}(GeV)^{-1}$. \\

Now, following a standard quantization procedure, the conmutator relation for $\delta\phi$ and its canonical conjugate momentum $\Pi^{0}_{(\delta\phi)}=\frac{\partial L}{\partial \dot{\delta\phi}}$ is given by
\begin{equation}\label{inffn1}
\left[\delta\phi(t,\bar{x}),\Pi^{0}_{(\delta\phi)}(t,\bar{x}^{\prime})\right]=i\delta^{(3)}(\bar{x}-\bar{x}^{\prime}).
\end{equation}
Thus, using  $\Pi^{0}_{(\delta\phi)}=\sqrt{-h}\,[(\dot{\phi}_c+\dot{\delta\phi})]$ the commutator (\ref{inffn1}) reads
\begin{equation}\label{inffn2}
\left[\delta\phi(t,\bar{x}),\dot{\delta\phi}(t,\bar{x}^{\prime})\right]=\frac{i}{\sqrt{-h}}\,\delta^{(3)}(\bar{x}-\bar{x}^{\prime}).
\end{equation}
We introduce the auxiliary field $\delta\chi$ as
\begin{equation}\label{yc1}
\delta\phi(t,\bar{x})=\exp\left(-\frac{3}{2}\int H(t)dt\right)\delta\chi(t,\bar{x}).
\end{equation}
We consider the Fourier expansion 
\begin{equation}\label{inff15}
\delta\chi(t,\bar{x})=\frac{1}{(2\pi)^{3/2}}\int d^{3}k\left[a_{k}e^{i\bar{k}\cdot\bar{x}}\eta_{k}(t)+a_{k}^{\dagger}e^{-i\bar{k}\cdot\bar{x}}\eta_{k}^{*}(t)\right],
\end{equation}
with the asterisk mark denoting complex conjugate and, $a_{k}$ and $a_{k}^{\dagger}$ being the annihilation and creation operators. These operators satisfy the commutator algebra
\begin{equation}\label{inff17}
\left[a_{k},a^{\dagger}_{k^{\prime}}\right]=i\delta^{(3)}(\bar{k}-\bar{k}^{\prime}),\qquad \left[a_{k},a_{k^{\prime}}\right]=\left[a^{\dagger}_{k},a^{\dagger}_{k^{\prime}}\right]=0.
\end{equation}
The quantum modes $\eta_k^{(end)}(t)$ at the end of inflation,  accor\-ding to (\ref{inff3}), \eqref{ulca4}, \eqref{sms4}, \eqref{sms6} and \eqref{yc1} are given by
\begin{equation}
\ddot{\eta}_k^{(end)}+\left[\frac{k^2}{\tilde{a}_e^2e^{2H_et}}-\frac{9}{4}H_e^2+U^{\prime\prime}(\phi_e)\right]\eta_k^{(end)}=0,
\label{inff16}
\end{equation}
where $H_e=H(t_e)$ and 
\begin{equation}\label{upp}
U^{\prime\prime}(\phi_e)=-\frac{\lambda\phi_e^2(-3+5\beta^2\phi_e^4)}{\xi^2(1+\beta^2\phi_e^4)}.
\end{equation}
Selecting the Bunch Davies condition \cite{BDV}, the normalized solution of \eqref{inff16} reads
\begin{equation}\label{yc2}
\eta_k^{(end)}=\frac{1}{2\tilde{a}_e}\sqrt{\frac{\pi}{\tilde{a}_e H_e}}{\cal H}_{\nu}^{(1)}[z(t)],
\end{equation}
where $\mathcal{H}_\nu^{(1)}$ is the first kind Hankel function, the parameter   $\nu=(1/2)\sqrt{9-(4U^{\prime\prime}(\phi_e)/H_e^2)}$ and $z(t)=[k/(\tilde{a}_e H_e)]e^{-H_et}$.\\

The amplitude of $\delta\phi$ on the infrared sector is given by the expression
\begin{equation}\label{inff19}
\left<\delta\phi^2\right>_{IR}=\frac{2^{2\nu}\Gamma^2(\nu)}{8\pi^3\tilde{a}_e^2}\frac{e^{-(3-2\nu)H_et_e}}{(\tilde{a}_e H_e)^{1-2\nu}} \int_{0}^{\varepsilon k_H}\frac{dk}{k}k^{3-2\nu},
\end{equation}
where $\varepsilon=k_{max}^{IR}/k_p\ll 1$ is a dimensionless parameter with $k_{max}^{IR}=k_{H}(t_r)$ being the wave number related to the Hubble radius at the time $t_r$, which is the time when the modes re-enter to the horizon and $k_p$ is the Planc\-kian wave number. It is well-known that for a Hubble parameter $H=0.5\times 10^{-9}\,M_p$, the values of $\varepsilon $ range between $10^{-5}$ and $10^{-8}$, and this corresponds to a number of e-foldings at the end of inflation $N_e=63$. Hence the squared $\delta\phi$-fluctuations has a power-spectrum
\begin{equation}\label{spectro}
\mathcal{P}_s(k)=\frac{2^{2\nu}\Gamma^2(\nu)}{8\pi^3\tilde{a}_e^2}\frac{e^{-(3-2\nu)H_et_e}}{(\tilde{a}_e H_e)^{1-2\nu}}k^{3-2\nu}.
\end{equation}
On the other hand, the scalar to tensor ratio $r$ and the scalar spectral index $n_s$ are given by  $r=16\epsilon$ and $1-n_s=6\epsilon-2\eta$, being $\epsilon$ and $\eta$ the slow-roll parameters
 \begin{equation}
 \epsilon=\frac{M_p^2}{2}\left(\frac{U^{\prime}}{U}\right),\quad \eta=M_p^2\left(\frac{U^{\prime\prime}}{U}\right).
 \end{equation}
 Thus, with the help of \eqref{ulca4} and \eqref{phiefol} the scalar spectral index is given by
 \begin{equation}\label{yc3}
 1-n_s=\frac{8\beta M_p^2\Delta(5\Delta^{4/3}-7\Delta^{2/3}+5)}{(\Delta^{2/3}-1)(\Delta^{4/3}-\Delta^{2/3}+1)^2}\simeq \frac{5}{3N}.
 \end{equation}
 It is not difficult to see from \eqref{yc3} that for a a number of e-foldings $N=63$ the scalar spectral index is appro\-ximately $n_s\simeq 0.9735$, which is in agreement with PLANCK observations: $n=0.968\pm0.006$ \cite{AR3}. 
Similarly, the scalar to tensor ratio results to be
\begin{equation}\label{yc4}
r=\frac{128M_p^2\beta\Delta^{5/3}}{(\Delta^{2/3}-1)(\Delta^{4/3}-\Delta^{2/3}+1)^2}\simeq\frac{128}{\beta^{2/3}M_p^{4/3}}\frac{1}{(24N)^{5/3}}.
\end{equation}
Again, it follows from \eqref{yc4} that for $N=63$ and the parameter $\beta\simeq 0.01629 M_p^{-2}$, that the scalar to tensor ratio is of the order $r\simeq 0.01$, in consistency with the PLANCK data ($r<0.11$). Hence, for such value of $\beta$ the parameter $\xi$ must ranges in the interval $(2.81047\cdot10^{7},2.81047\cdot10^{8})$. Thus, taking this value of $\xi$ the equation \eqref{upp} indicates that $U^{\prime\prime}(\phi_e)\ll1$, and therefore $\nu\simeq 3/2$ which accor\-ding to \eqref{spectro}  corresponds to a nearly scale invariant power spectrum at the end of inflation ${\cal P}_{s}(k)\sim H_{e}^2$.
\section{Final Remarks}

In this letter we have derived a Higgs inflationary model 
on the framework of a new kind of complex scalar-tensor theory of gravity that we called: geome\-tri\-cal complex  scalar-tensor theory of gravity. In this approach we consider a complex scalar-tensor theory compatible with it's background geometry. We mean by compatibility the invariance of the action under the symmetry group of its background geometry. The former is obtained here by employing the Palatini's variational  principle, resulting that the background geometry for a complex scalar-tensor theory is a kind of complex Weyl-integrable geometry. As has ocurred in the case of real geometrical scalar-tensor theories, there are two frames: the Weyl and the Riemann frames. The Riemann frame is obtained by a particular gauge election of the Weyl-transformations: $f=-\varphi$. In the Weyl frame the scalar field is part of the affine structure of the space-time manifold, whereas in  the Riemann frame it can be considered as a physical field. This is why general relativity can be recovered in the Riemann frame. \\

As an application of the formalism we developed a Higgs inflationary model. An interesting feature of our model is that the inflaton and the Higgs fields can be both identified with the Weyl scalar field. Moreover, due to the compatibility of the new complex scalar-tensor theory with its background geometry, the Higgs potential can be rescaled enough to generate the primordial inflation of the universe. We obtain a super-De-Sitter expansion at the beginning of inflation. The infrared power spectrum results nearly scale invariant at the end of inflation for $\beta\simeq 0.01629\,M_p^{-2}$. For $N=63$ e-foldings we obtain an spectral index $n_s\simeq 0.9735$ and a scalar to tensor ratio $r\simeq 0.01$, which are in agreement with PLANCK observations \cite{AR3}.

\section*{Acknowledgements}

\noindent  J.E.Madriz-Aguilar, J. Zamarripa and M. Montes  acknowledge CONACYT
M\'exico, Centro Universitario de Ciencias Exactas e Ingenierias and Centro Universitario de los Valles of Universidad de Guadalajara for financial support. C. Romero thanks Cnpq for partial financial help

\end{document}